\documentclass[conference]{IEEEtran}
\usepackage{cite}
\usepackage{amsmath,amssymb,amsfonts}
\usepackage{algorithmic}
\usepackage{graphicx}
\usepackage{textcomp}

\usepackage{xcolor}
\def\BibTeX{{\rm B\kern-.05em{\sc i\kern-.025em b}\kern-.08em
    T\kern-.1667em\lower.7ex\hbox{E}\kern-.125emX}}

\usepackage{hyperref}

\newcommand{\comment}[1]{}

\newcommand{\brackref}[1]{\hyperref[#1]{(\autoref*{#1})}}

\usepackage{subcaption}

\usepackage[all]{nowidow}

\usepackage{makecell}

\usepackage{array}
\newcolumntype{P}[1]{>{\centering\arraybackslash}p{#1}}

\usepackage{float}

\usepackage{booktabs}

\usepackage{siunitx}

\usepackage{color}
\definecolor{deepblue}{rgb}{0,0,0.5}
\definecolor{deepred}{rgb}{0.6,0,0}
\definecolor{deepgreen}{rgb}{0,0.5,0}

\usepackage{listings}
\usepackage{etoolbox}
\newcounter{BalanceAtReference}
\newcounter{ReferenceIndexForBalancing}
\makeatletter
\def\@balancelastpageonce{%
	\ifnum\value{ReferenceIndexForBalancing}=\value{BalanceAtReference}
	\newpage
	\else
	\relax
	\fi
	\stepcounter{ReferenceIndexForBalancing}
}
\pretocmd{\bibitem}{\@balancelastpageonce}
{} 
{\@latex@error{Patching \bibitem failed}{\@ehd}}
\makeatother

\begin{document}

\title{Jaco: An Offline Running Privacy-aware \mbox{Voice Assistant}}

\comment{
	\author{\IEEEauthorblockN{1\textsuperscript{st} Anonymous Researcher}
		\IEEEauthorblockA{\textit{Unknown Institute} \\
			\textit{Secret University}\\
			Earth, Solar System \\
			a.r.1@internet.com}
		\and
		\IEEEauthorblockN{2\textsuperscript{nd} Anonymous Researcher}
		\IEEEauthorblockA{\textit{Unknown Institute} \\
			\textit{Secret University}\\
			Earth, Solar System \\
			a.r.2@internet.com}
		\and
		\and
		\IEEEauthorblockN{3\textsuperscript{rd} Anonymous Researcher}
		\IEEEauthorblockA{\textit{Unknown Institute} \\
			\textit{Secret University}\\
			Earth, Solar System \\
			a.r.3@internet.com}
	}
}

\author{\IEEEauthorblockN{Daniel Bermuth}
	\IEEEauthorblockA{\textit{ISSE} \\
		\textit{University of Augsburg}\\
		Augsburg, Germany \\
		daniel.bermuth@informatik.uni-augsburg.de}
	\and
	\IEEEauthorblockN{Alexander Poeppel}
	\IEEEauthorblockA{\textit{ISSE} \\
		\textit{University of Augsburg}\\
		Augsburg, Germany \\
		poeppel@isse.de}
	\and
	\and
	\IEEEauthorblockN{Wolfgang Reif}
	\IEEEauthorblockA{\textit{ISSE} \\
		\textit{University of Augsburg}\\
		Augsburg, Germany \\
		reif@isse.de}
}

\comment{
	\author{\IEEEauthorblockN{1\textsuperscript{st} Daniel Bermuth \hspace{28pt} 2\textsuperscript{nd} Alexander Poeppel \hspace{28pt} 3\textsuperscript{rd} Wolfgang Reif}
		\IEEEauthorblockA{\textit{Institute for Software \& Systems Engineering} \\
			\textit{University of Augsburg}\\
			Augsburg, Germany \\
			\{daniel.bermuth,alexander.poeppel,reif\}@informatik.uni-augsburg.de}
	}
}

\maketitle

\begin{abstract}
	With the recent advance in speech technology, smart voice assistants have been improved and are now used by many people. But often these assistants are running online as a cloud service and are not always known for a good protection of users' privacy. 
	This paper presents the architecture of a novel voice assistant, called \textit{Jaco}, with the following features:
	(a)~It can run completely offline, even on low resource devices like a RaspberryPi.
	(b)~Through a skill concept it can be easily extended.
	(c)~The architectural focus is on protecting users' privacy, but without restricting capabilities for developers.
	(d)~It supports multiple languages.
	(e)~It is competitive with other voice assistant solutions.
	In this respect the assistant combines and extends the advantages of other approaches.
\end{abstract}

\begin{IEEEkeywords}
multilingual smart voice assistant; human-computer interaction; offline voice assistant
\end{IEEEkeywords}

\section{Introduction}
\label{sec:intro}

Smart voice assistants have greatly improved over the last years and are now helping in various tasks in many households. Often these assistants are running online as a cloud service and are not always known for a good protection of users' privacy. On the other hand,  some approaches exist which try to improve the privacy aspect. One important concept in this regard is that the assistant can run completely offline. In typical cloud solutions users have no guarantee that the voice commands sent to online servers are handled safely there. The only option would be to fully trust the cloud provider.

\vspace{9pt}
Amazon's \textit{Alexa}~\cite{ALXA} is a widely used voice assistant, which can understand multiple languages. It is normally shipped with specialized hardware (the \textit{Echo} speakers) and offers a large skill store with both free and commercial skills.
\textit{Snips}~\cite{SNIPS} was a voice assistant, capable of running offline, and still achieving a high recognition accuracy. It had a small skill store where hobby developers could share their skills. The company behind Snips was bought by Sonos and the possibility to create own assistants was removed.
\textit{Mycroft}~\cite{MYCRF} is a fully open source assistant and has, like \textit{Snips}, a focus on preserving privacy. For speech recognition \textit{Mycroft} uses Google's speech-to-text service.
\textit{Rhasspy}~\cite{RHSPY} combines different services into a voice assistant capable of running offline. The project is focused on creating a voice interface for home automation software like \textit{Home Assistant}. Unlike the other alternatives it does not have a skill store for which developers can build specialized skills in order to share them with others.

Besides the above mentioned assistants, several solutions exist that are specialized on the speech to intent extraction, which is the most important part of an assistant, regarding its command recognition performance. Google's \textit{DialogFlow}, Microsofts's \textit{LUIS} and IBM's \textit{Watson} are cloud-based solutions and Picovoice offers an offline running alternative with \textit{Rhino}~\cite{RHNO}.

\vspace{9pt}
This paper presents a novel voice assistant that combines the benefits of the aforementioned approaches. The main advantages of \textit{Jaco} are:
(a)~It can run completely offline, in contrast to \textit{Snips} where only the usage was offline, but the training of the assistant had to be done online, similar to the cloud-based training of \textit{Rhino}. \textit{Mycroft} uses the online speech recognition service from Google, and \textit{Alexa} runs completely in the cloud.
(b)~By adding skills to the assistant it can be easily extended with new features or integrated into other frameworks like the \textit{Robot Operating System}~\cite{ROS}.
(c)~While keeping architectural focus on protecting users' privacy, the developers are not restricted in using all of the host device's resources.
(d)~It supports multiple languages, which currently are German, English, Spanish and French, and can easily be extended with new ones.
(e)~It is competitive with other voice assistant solutions and outperforms them in various benchmarks.

The complete assistant and the benchmark code can be found at: https://gitlab.com/Jaco-Assistant

\section{Architecture}
\label{sec:arch}

\begin{figure}[htb]
	\centering
	\includegraphics[width=0.95\linewidth]{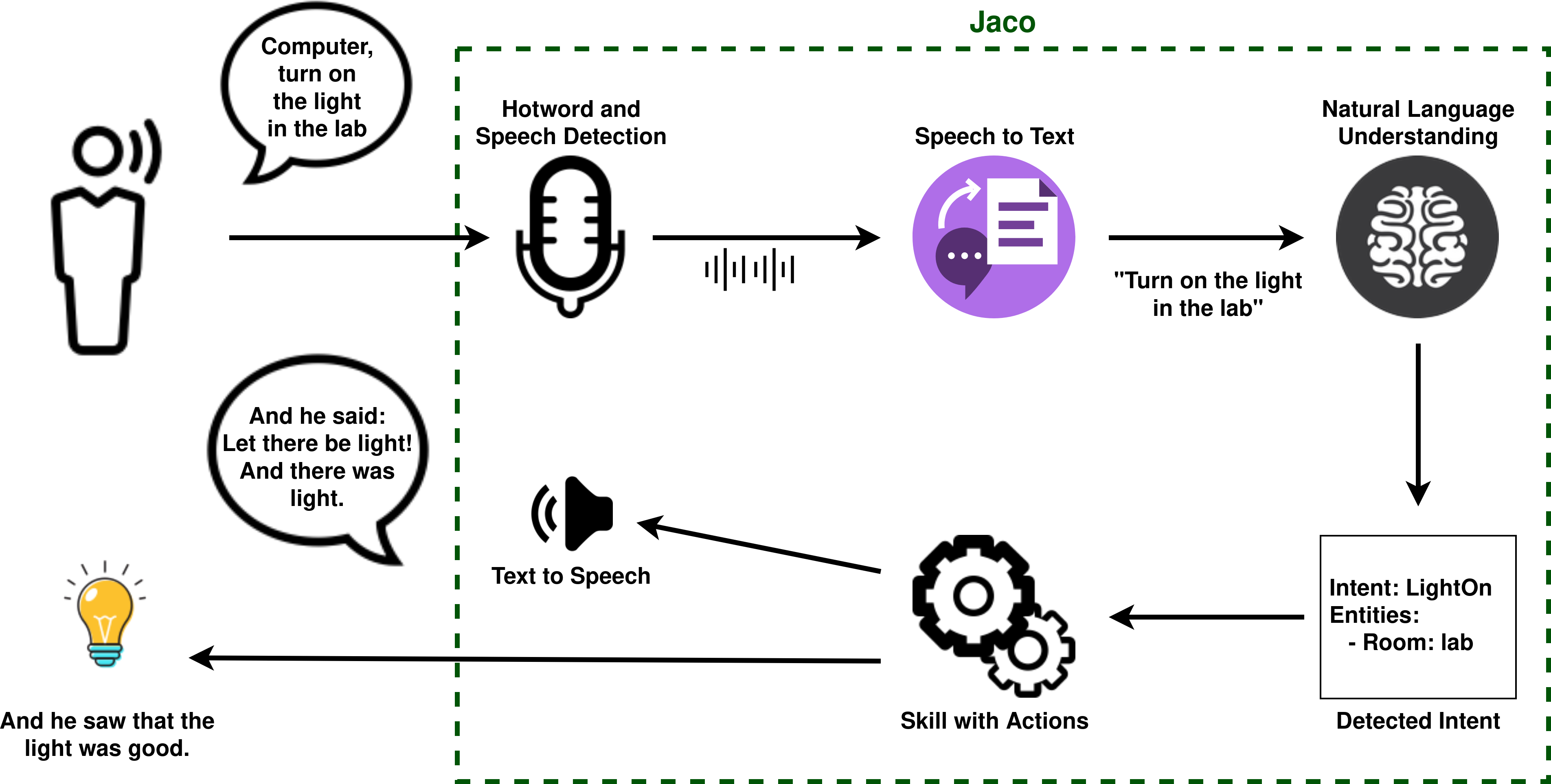}
	\caption{General architecture of a voice assistant.}
	\label{fig:jflow}
\end{figure}

The general architecture of a typical voice assistant is shown in Figure~\ref{fig:jflow}. Usually an interaction starts with a user speaking a \textit{keyword} (here: \textit{computer}) that triggers the assistant to listen to the following spoken request. A \textit{Speech-to-Text} module transcribes the request and the transcription is sent to a \textit{Natural-Language-Understanding} module, which extracts the useful information from the full sentence. The extracted intent and the entities are then handled by a skill, which performs the appropriate action and informs the user upon success, by sending a textual answer to a \textit{Text-to-Speech} service, which answers the user.

\begin{figure}[htb]
	\centering
	\includegraphics[width=\linewidth]{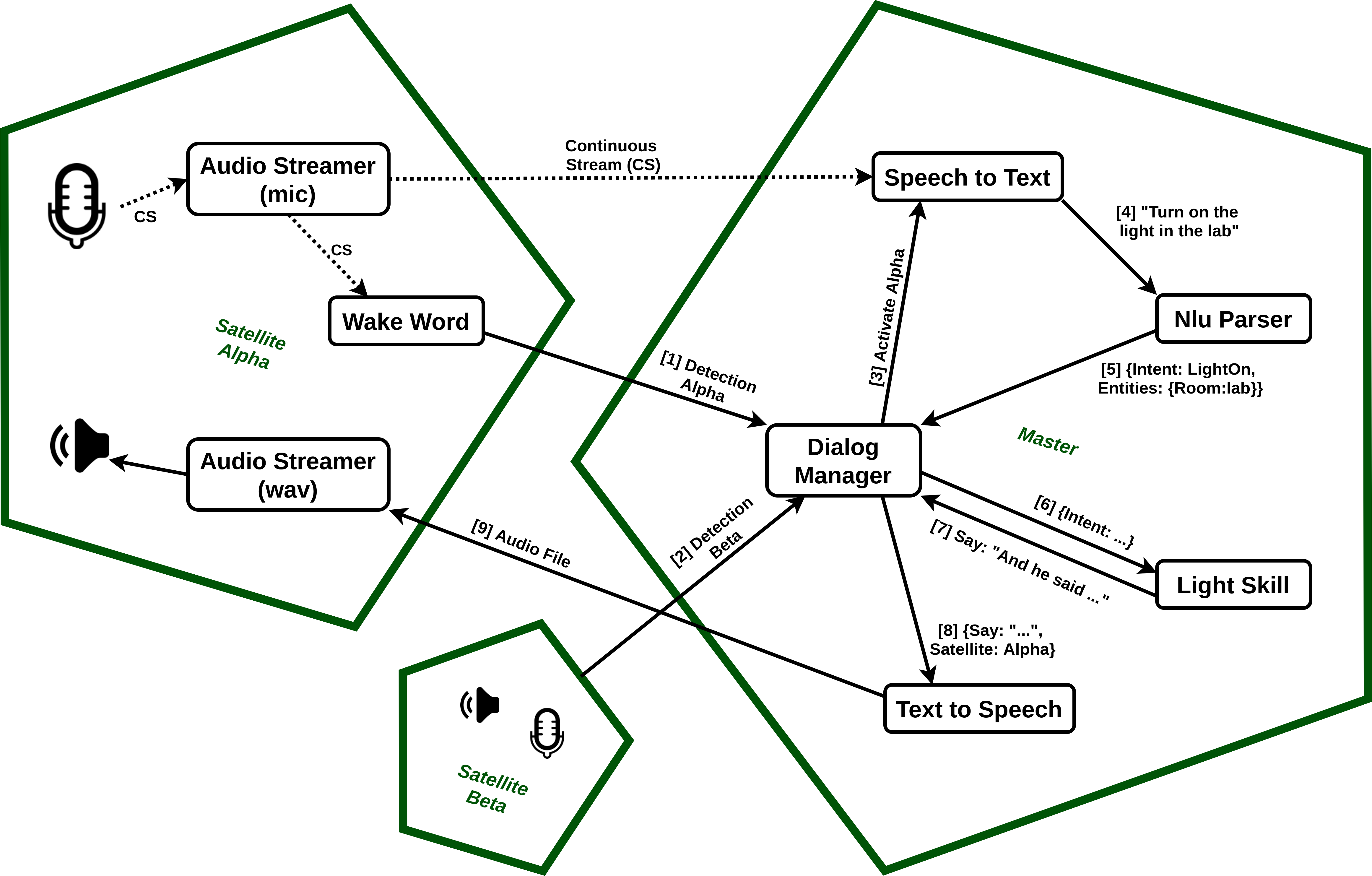}
	\caption{Process of a voice interaction using Jaco.}
	\label{fig:jcomm}
\end{figure}

\vspace{9pt}
The architecture of the proposed assistant \textit{Jaco} is presented in greater detail in Figure~\ref{fig:jcomm}. The general architecture is split into two parts, one or multiple lightweight Satellites, which can be installed in different rooms, and a single central node, which handles the main processing tasks.
The example consists of two different Satellites (\textit{Alpha}, \textit{Beta}) and the \textit{Master}.
Both satellites have a microphone from which the \textit{Audio Streamer} reads the audio input
and streams it to the \textit{Wake Word} modules. If they detect the wake word in the stream,
they send a message to the \textit{Dialog Manager}. 
After checking which Satellite detected the wake word first,
the \textit{Dialog Manager} sends an activation request to the \textit{Speech to Text} module.
The transcription of the speech is then started and continues until the user stops speaking.
Afterwards the detected text is sent to the \textit{Nlu Parser} which extracts the intent
and sends it back to the \textit{Dialog Manager}. From there it is sent to the skill
which handles the detected intent type. 
The \textit{skill} can run various actions (like turning on the light in the lab)
and responds with a text message. The \textit{Dialog Manager} then forwards the message to the \textit{Text to Speech} module. There the text is converted
into an audio file which afterwards is sent to the Satellite specified in the message.
The \textit{Audio Streamer} then plays the file to the user through the speaker.

\vspace{9pt}
All the modules are container-based and run in their own individual container, which has the advantage of a reduced potential of conflicts with programs already installed on the host system. All module requirements can also be preinstalled into the containers, which greatly improves the installation time, as users can download usage-ready container images, instead of building them on their own device (which would take about \SI{12}{\hour} on a RaspberryPi-4).

For the wake-word recognition \textit{Porcupine}~\cite{PORCU} is used. The speech-to-text module uses the pretrained models from \textit{Scribosermo}~\cite{SCRSO}, which are designed to run in real-time on a RaspberryPi. \textit{Scribosermo} was developed for usage with \textit{Jaco} and the source code can be found in the assistant's repository group. The NLU-parsing is done with \textit{Rasa}~\cite{RASA} and \textit{Duckling}~\cite{DUCKL}, and the text-to-speech generation uses \textit{picotts}~\cite{PITTS}. Communication between the modules is executed over MQTT.
The solutions were selected because they can run completely offline on a RaspberryPi. An advantage of the module based approach of \textit{Jaco} is that the assistant doesn't rely on specific solutions, so they can easily be replaced by any other service. This for example allows other researchers to replace specific modules with their current research, with the benefit that they can test it in a complete workflow.

Two of the modules, the \textit{Nlu Parser} and the \textit{Speech to Text} module, require additional training, depending on the skills installed by a user. For the speech recognition, the acoustic model, which maps an audio input to character possibilities, differs only between different languages, but a customized \textit{n-gram} language model (LM) is created, which rescores the character predictions, using the example requests included in the skill files. The NLU model also has to be retrained on the skill's example requests. In both cases a script collects the command examples from the installed skills and extends them with automatically generated examples. This is done by inserting different entity values into the example intents. Training both models usually takes a few seconds on a computer and a few minutes on a RaspberryPi, but this depends on the installed skills. Adding a new language would only require a general purpose STT model for this language, as well as a TTS model, the rest of the system is language independent.

\section{Skills}
\label{sec:skills}

One of the most important features of a smart voice assistant is the possibility to add new capabilities in form of user created skills. The architecture of \textit{Jaco} was designed in a way that this is very easy to achieve, and, in contrast to the other assistants, the skills can also access all the hardware resources of the executing device.

To create a skill, a developer first has to define possible user requests. The syntax for the dialog examples is designed in a way that makes them easily readable when they are inspected in the skill's git repository. In the following example, the sample request would only be displayed as \textit{``Book (me$\vert$us) a flight from Augsburg to Berlin''}, with the two cities highlighted as links, which are referring to the \textit{city.txt} file. The roles \textit{start} and \textit{destination}, which are needed for a later distinction of the two cities in the action code, are separated with a question mark so that the names of the roles are ignored if a user clicks on the link. Words in parentheses define alternatives or synonyms. About ten example sentences per intent is a good amount to start with, but using more can slightly improve accuracy.

\noindent
\begin{minipage}{\linewidth}
\begin{lstlisting}[keywordstyle=\color{black}]
========================== city.txt ========================
Augsburg
(New York|N Y)->New York
Berlin	

=========================== nlu.md =========================
## lookup:city
city.txt

## intent:book_flight
- Book (me|us) a flight from [Augsburg](city.txt?start) \
     to [Berlin](city.txt?destination)
\end{lstlisting}
\end{minipage}

\noindent
Besides that, the developer also has to create a python script which can handle the incoming requests. To simplify development, all specialized message interactions are handled by the \textit{jacolib} library. After creating a skill it can be shared with other users, by adding it to a skill store.

\noindent
\begin{minipage}{\linewidth}
\begin{lstlisting}
========================= action.py ========================
from jacolib import assistant
assist = assistant.Assistant()

def callback_book(msg):
  locs = assist.extract_entities(msg, "myskill-book_flight")
  locs = [lc["value"] for lc in locs]                      
  if "munich" in locs:
    r = "that wouldn't be wise"
  else:
    r = "ok boss"
  assist.publish_answer(r, msg["satellite"])

assist.add_topic_callback("book_flight", callback_book)
assist.run()
\end{lstlisting}
\end{minipage}

\section{Improving privacy}
\label{sec:impry}

As mentioned before, the most important aspect regarding privacy is that the complete assistant can run entirely offline.
In contrast to \textit{Snips}, which required training the assistant on an online server before it could be downloaded, the training is executed completely offline on device as well. Running everything on the user's local device ensures that users don't have to trust their cloud provider that voice commands or accidentally recorded sounds (if the wake-word was triggered through a false positive) are handled safely there.

\vspace{9pt}
Besides that, multiple features were implemented to protect privacy when third-party skills are used, without restricting the possible use-cases too much.
An important part of this is that all skills must include a configuration file which implements a simple permission system.
In the config file all communication topics the skill wants to access have to be listed. By design all MQTT-topics are automatically encrypted to ensure that a skill can not read topics of other skills or the main modules. With the listing in the config file a user can easily see which information the skill wants to read, without restricting a skill's capabilities if they require reading other topics. For example, a weather skill should normally not require to read the microphone recording stream, but a music playing skill might use this data to automatically adjust the playback volume to background noises.
In the config file the developer also has to state if the skill needs internet access and if it runs an action. An example of a skill without an action is one that contains only specific dialog examples for sharing them with other skill developers.

\noindent
\begin{minipage}{\linewidth}
\begin{lstlisting}
========================= config.yaml ======================
system:
  has_action: true
  extra_container_flags: ""
  needs_internet_access: true
  topics_read:
    - "book_flight"
  topics_write:
    - "Jaco/Skills/SayText"
\end{lstlisting}
\end{minipage}

\vspace{9pt}
Unlike other assistants, where skills can not access the hardware resources of the executing device directly (Alexa) or require extra setup steps (Snips and Mycroft), skills for \textit{Jaco} can fully access the device resources. To improve the security of this feature, all skills are executed inside containers. This also allows the automatic installation of arbitrary additional software without the problem of creating software conflicts on the host device. 
In some cases the containers require additional runtime flags to access specific resources, which have to be added in the configuration, too.

An example would be a skill that wants to control \mbox{GPIO-pins} on a Raspberry\,Pi. With \textit{Snips} this could be achieved too, but the user had to install all required system libraries (like \textit{WiringPi}) and do the setup himself as the skill could only install new python libraries. With Jaco the skill developer can automate this through the skill container which then is granted access to the necessary resources. Another example could be a skill that works as interface for \textit{ROS} where the skill runs the multi-step installation automatically in its container. By allowing this skill to listen to system topics, the interface can optionally be used to pause robot movements to reduce disturbing noises while the user or system is speaking. A demo for both skills can be found in the skill store.

\begin{figure}[htb]
	\centering
	\includegraphics[width=0.91\linewidth]{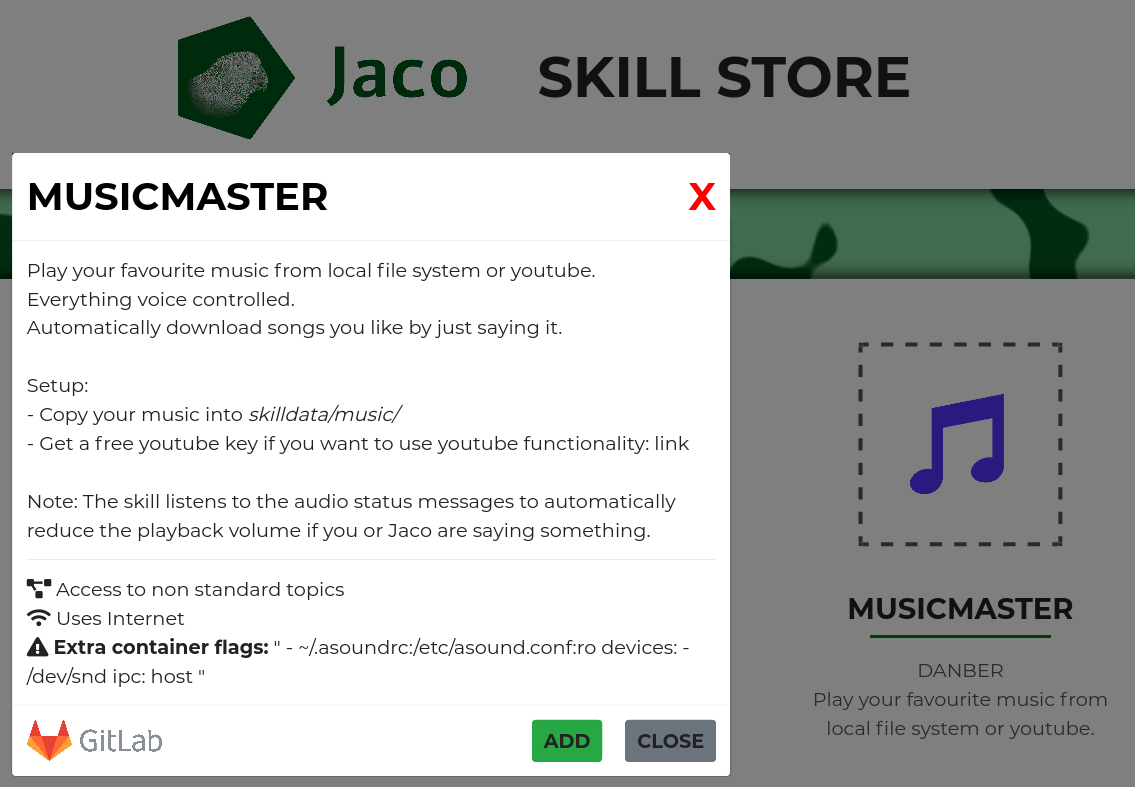}
	\caption{Store with permission notifications of the selected skill.}
	\label{fig:jstor}
\end{figure}

\vspace{9pt}
If users want to install shared third-party skills from the official skill store, they can directly see if skills request possibly dangerous access permissions (Figure~\ref{fig:jstor}).

The store entry also includes a link to the git repository from where the skill is downloaded later, so that a user can easily inspect the source code. Using the official store is not mandatory, a skill's code also can be shared through any other provider. 
The store can run completely offline in the browser after opening the website or can be rebuilt and executed from the source release.

\section{Benchmarks}
\label{sec:benchs}

\begin{figure}[H]
	\centering
	\includegraphics[width=0.88\linewidth]{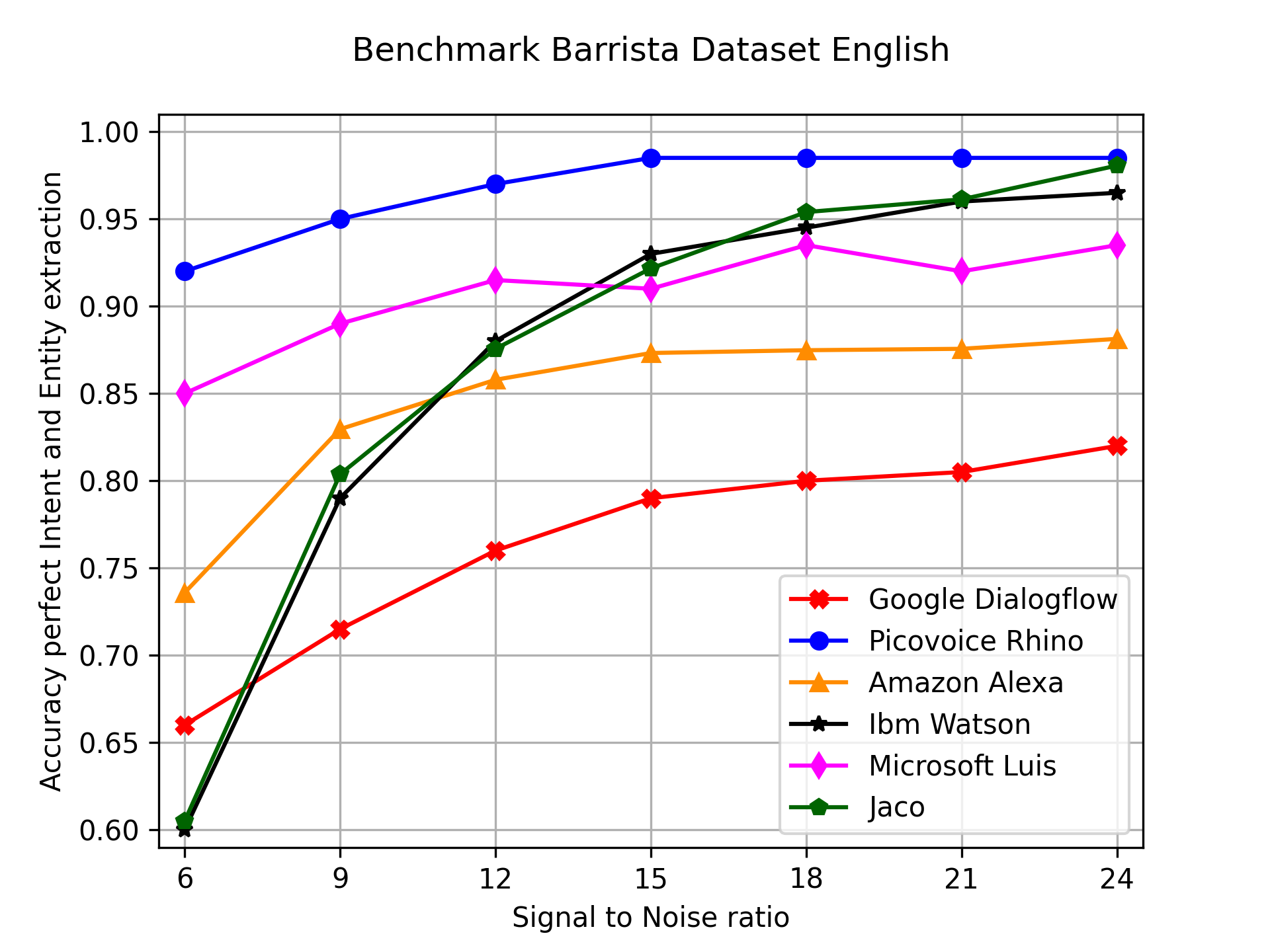}
	\caption{Benchmark coffee orders with noisy backgrounds. The results of \textit{DialogFlow}, \textit{Watson}, \textit{Luis} and \textit{Rhino} have been taken from~\cite{RHIBEN}.}
	\label{fig:bb}
\end{figure}

To compare the performance relative to other Speech and Language Understanding (SLU) solutions, different benchmarks were performed. The first benchmark was published by \textit{Picovoice}~\cite{RHIBEN} and consists of 620 commands of different people ordering coffee in English. The audio is mixed with different volume levels of background noise from cafe and kitchen environments. An example command would be: \textit{``i'd like a [medium roast] [large] [mocha] with [lots of cream] and [a little bit of brown sugar]"}. A command is correctly detected if the intent, as well as all the slots, could be retrieved by the assistant. The results are shown in Figure~\ref{fig:bb}. The benchmark shows that \textit{Jaco} outperforms most other solutions in medium and low noise settings.

\vspace{9pt}
Another often used capability of smart voice assistants is controlling lights in different rooms. This was tested with the \textit{SmartLights} benchmark from \textit{Snips}~\cite{SNSLU}. It consists of 1660 requests which are split into five partitions for a 5-fold training of the LM and NLU components. A sample command could be: \textit{``please change the [bedroom] lights to [red]"} or \textit{``i'd like the [living room] lights to be at [twelve] percent"}. \textit{Jaco} could outperform all the other comparable solutions (Table~\ref{tab:ben_slc}). 

Some notes on the execution: For \textit{Alexa}, a new skill had to be created for each of the five folds. \textit{Houndify}~\cite{HOUND} was tested because it claims \textit{``to deliver unprecedented speed and accuracy"} but did not provide any quantifiable evidence for this claim. However it was only possible to use the pretrained domain for smart light controls. Commands such as \textit{``it's too dark in here"}, which should trigger the \textit{SwitchLightOn} intent, were often not understood. For a better comparison the Word-Error-Rate (WER) was measured as well. \textit{Houndify} and \textit{Jaco} both had a WER of $10.8\%$, for \textit{Alexa} no method for returning the transcriptions was found.

\begin{table}[htb]
	\caption{Benchmark smart light control commands.}
	\label{tab:ben_slc}
	\centering
	\begin{tabular}{lcc}
		\toprule
		\textbf{} & \textbf{Accuracy} & \textbf{WER} \\
		\midrule
		\textit{Google}~\cite{SNSLU} & $0.793$ & $-$ \\
		\textit{Snips}~\cite{SNSLU} & $0.842$ & $-$ \\
		\textit{Alexa} & $0.792$ & $-$ \\
		\textit{Houndify} & $0.545$ & $0.108$ \\
		\textit{Jaco} & $\textbf{0.854}$ & $0.108$ \\
		\textit{AT-AT}~\cite{ATAT} & $0.849$ & $-$ \\
		\textit{SynSLU}~\cite{SYNSLU} & $0.714$ & $-$ \\
		\bottomrule
	\end{tabular}
\end{table}

\vspace{9pt}
The last benchmark (Table~\ref{tab:ben_ssc}) tests the performance of reacting to music player commands in English as well as in French. The benchmark is from \textit{Snips}~\cite{SNSLU}, too, and is the only one that could be found that includes a language other than English. It has the difficulty of containing many artist or music tracks with uncommon names in the commands, like \textit{``play music by [a boogie wit da hoodie]"} or \textit{``I'd like to listen to [Kinokoteikoku]"}. In the English benchmark none of the solutions performed very well. The authors of the \textit{Snips} benchmark did precompute pronunciation mappings for the artist names, which was not done for \textit{Jaco}. While benchmarking \textit{Alexa} it could be noticed that stylized artist names like \textit{"Bonez MC"} often were automatically matched to the spelling of the originating words (\textit{"Bones MC"}) and are therefore classified as incorrect slot extractions.

In the French benchmark \textit{Alexa} performed surprisingly well, the spelling correction issue did not occur anymore. \textit{Snips} performed well, too, the generated pronunciations were automatically mapped to a French spelling. \textit{Jaco} had great problems with recognizing many of the artists' names. Something like a pronunciation map for the names could help here, but was not implemented. But it is generally possible to ship such a pronunciation map with a skill, if the skill author chooses to create one.

\begin{table}[htb]
	\caption{Benchmark music wishes in English and French.}
	\label{tab:ben_ssc}
	\centering
	\begin{tabular}{lcc}
		\toprule
		\textbf{Accurracy:} & \textbf{English} & \textbf{French} \\
		\midrule
		\textit{Snips}~\cite{SNSLU} & $\textbf{0.687}$ & $0.751$ \\
		\textit{Google}~\cite{SNSLU} & $0.478$ & $0.423$ \\
		\textit{Jaco} & $0.627$ & $0.480$ \\
		\textit{Alexa} & $0.455$ & $\textbf{0.889}$ \\
		\bottomrule
	\end{tabular}
\end{table}

\section{Conclusion}
\label{sec:conclu}

In this paper the full-featured voice assistant \textit{Jaco} was presented. The assistant combines and extends the advantages of other approaches. It is competitive with other solutions and can run on single-board computers like a RaspberryPi. The assistant has a strong focus on protecting users' privacy, and runs completely offline, so no user interactions are shared with any other service.

\bibliographystyle{IEEEtran}
\bibliography{mybib}

\end{document}